\documentclass[fleqn,usenatbib]{mnras}

\usepackage{newtxtext,newtxmath}

\usepackage[T1]{fontenc}
\usepackage{ae,aecompl}

\usepackage{graphicx}	
\usepackage{amsmath}	
\usepackage{changes}
\usepackage{relsize}
\usepackage{adjustbox}

\def\aj{AJ}
\def\apj{ApJ}
\def\apjl{ApJL}
\def\apjs{ApJ Suppl. Ser.}

\def\aap{A\&A}

\def\beq#1{\begin{equation}\label{#1}}
\def\eeq{\end{equation}}
\def\beqa#1{\begin{eqnarray}\label{#1}}
\def\eeqa{\end{eqnarray}}

\def\comment#1{\relax}

\newcommand{\ZTF}{ZTF J0538+1953 }

\title[Double WD \ZTF]{Double white dwarf ZTF J0538+1953 as the brightest verification binary for space laser interferometers}

\author[S.V. Antipin et al.]{
S.V. Antipin$^1$\thanks{E-mail: serge\_ant@inbox.ru},
A.A. Belinski$^1$,
L.N. Berdnikov$^1$,
A.M. Zubareva$^{2,1}$,\newauthor
N.A. Maslennikova$^1$,
K.A. Postnov$^1$\thanks{E-mail: kpostnov@gmail.com, the corresponding author} and 
I.A. Strakhov$^1$
\\
$^{1}$ Sternberg Astronomical Institute, 13, Universitetskij pr., 119234, Moscow, Russia\\
$^{2}$ Institute of Astronomy RAS, 48 Pyatnitskaya str., Moscow, Russia
}

\date{To appear in Astronomy Letters}

\pubyear{2025}

\begin{document}
\label{firstpage}
\pagerange{\pageref{firstpage}--\pageref{lastpage}}
\maketitle

\begin{abstract}
A decrease in the orbital period of the ultrashort-period binary  white dwarf \ZTF, which is one of the Galactic verification binaries in the millihertz frequency range for planned space laser interferometers, has been measured.
Based on photometric observations carried out on the 2.5-m telescope of the Caucasian Mountain Observatory of the Sternberg Astronomical Institute of Moscow State University (CMO SAI MSU), a diagram \textit{O-C} is constructed. It can be described by quadratic elements of the brightness variation, which correspond to a decrease rate of the orbital period of the system of $dP/dt=-(1.16\pm 0.22)\times 10^{-11}$ s/s. The decrease rate of the orbital period in the quadrupole approximation for the emission of gravitational waves by a binary system corresponds to its chirp mass $\mathcal{M}=0.434\pm 0.05 M_\odot$, which turned out to be $\sim 30\%$ higher than the value obtained earlier from spectroscopic mass determination.
The chirp mass of \ZTF inferred from the measured orbital decay rate  makes this system the brightest Galactic verification binary for LISA and TianQin space interferometers with a signal-to-noise ratio of $\approx 119$ and $\approx 30$ over 5 years and 2.5 years of observations, respectively.

\end{abstract}

\begin{keywords}
close binary stars, white dwarfs, gravitational waves
\end{keywords}



\section{Introduction}
Currently, astronomy is undergoing a transition to the era of multimessenger observations, which include not only electromagnetic, but also gravitational wave and neutrino radiation. On the eve of the launch of the LISA, TaiJi and TianQin millihertz space gravitational wave interferometers
\citep{2024arXiv240207571C,2021PTEP.2021eA108L,2025CQGra..42q3001L}, the study of a class of Galactic sources potentially detected by these observatories, ultraclose  binary systems, is of key importance. These binaries include, in particular, close double white dwarfs with orbital periods of less than one hour, the evolution of which is mainly driven by the energy losses to gravitational radiation.


Ultraclose binary systems play a critical instrumental role for the success of planned space missions. Such binaries with well-defined orbital parameters (orbital period, component masses, orbital inclination, etc.) and distance are considered as "verification binaries" -- the main sources for calibrating space laser interferometers in order to test their accuracy. In addition, the combined radiation from millions of close binary systems of the Milky Way forms the Galactic GW background which should be taken into account in searching for other GW signals.


In 2023, we started photometric monitoring of the verification eclipsing binary systems from the list \citep{2023ApJS..264...39R}, 
available for observations by telescopes of SAI MSU, in order to determine their orbital period decay due to the GW radiation. The results for the ZTF J213056.71+442046.5 were published in \citep{2023PZ.....43...10A,2024AstL...50..619A}. For this ultraclose semi-detached binary system with a period of $P=0^d.0273195154$, consisting of a white dwarf and a low-mass hot helium sub-dwarf \citep{2020ApJ...891...45K}, the Hertzsprung method was applied to construct an \textit{O-C} diagram for the moments of the eclipse minima. The analysis of this diagram revealed a decrease in the orbital period at a rate of $dP/dt=-(2.66\pm 0.62)\times 10^{-12}$ s/s.


This value turned out to be within the errors consistent with the value predicted by the quadrupole formula for spectroscopically determined masses of the components of ZTF J213056.71+442046.5 \citep{2020ApJ...891...45K}. However, the value of the chirp mass ${\cal{M}}=\frac{(M_1M_2)^{3/5}}{(M_1+M_2)^{1/5}}$ derived from the measured orbital period decay rate turned out to be slightly higher than that calculated from spectroscopically determined masses, which increases the expected signal to noise ratio of the GW signal from this object. When analyzing  ZTF J213056.71+442046.5, we used a 6.5-year observation interval, including the first data from the ZTF survey and our own observations of the system obtained with the RC600 telescope of CMO SAI MSU \citep{2020ARep...64..310B}.


In this paper, we present the results of observations of an ultrashort-period binary eclipsing binary white dwarf\ZTF with an orbital period of $P=0^d.010030116$ (14.44 min) \citep{2020ApJ...905...32B}. Due to the faintness of the object ($R_c\approx 19^m.25 - 20^m.$2) the photometry was carried out on the 2.5-m telescope of CMO SAI MSU during 2024-2025. The quality of the ZTF photometry for this system turned out to be insufficient for use in analyzing the variability of the orbital period.

\section{Observations and data processing}


Photometric observations of $\ZTF$ ($\alpha = 05^h 38^m 02^s.724$, $\delta = +19\circ 53^\prime 02^{\prime\prime}.98$ (J2000), GAIA DR3) were carried out by us for eight nights from April 2024 to October 2025 on the 2.5-m telescope of CMO SAI MSU \citep{2017ARep...61..715P}, equipped with a 4k$\times$4k CCD camera, detailed information about which will be provided in the next section. The duration of the monitoring during one night was at least 1.5 hours. A total of 5,686 CCD frames were obtained with an exposure of 10 seconds in the $R_c$ band with two binning and four-channel reading. The observation log is presented in Table~\ref{tab:log}. All time moments were reduced to the Solar System barycenter, which is necessary to search for variations in the orbital period of ultrashort-period eclipsing binaries.

When reading into one channel, the bias was taken into account  by overscan, i.e. reading additional pixels at the edges of the frame (which do not physically exist) by adding additional reading cycles. The line-by-line median in the overscan was calculated, then it was line-by-line subtracted from the scientific frame.


When reading into four channels, there is no overscan area, therefore, 5-10 dark current frames with the same exposure as scientific frames were obtained immediately after the series. Next, the pixel-by-pixel median of these dark frames was subtracted from the scientific frames. The flat field frames were taken in series of 6 pieces across the twilight sky in the required filters and reading modes. The scientific frames were divided into a normalized flat field.

\begin{figure}
    \centering
    \includegraphics[width=0.9\linewidth]{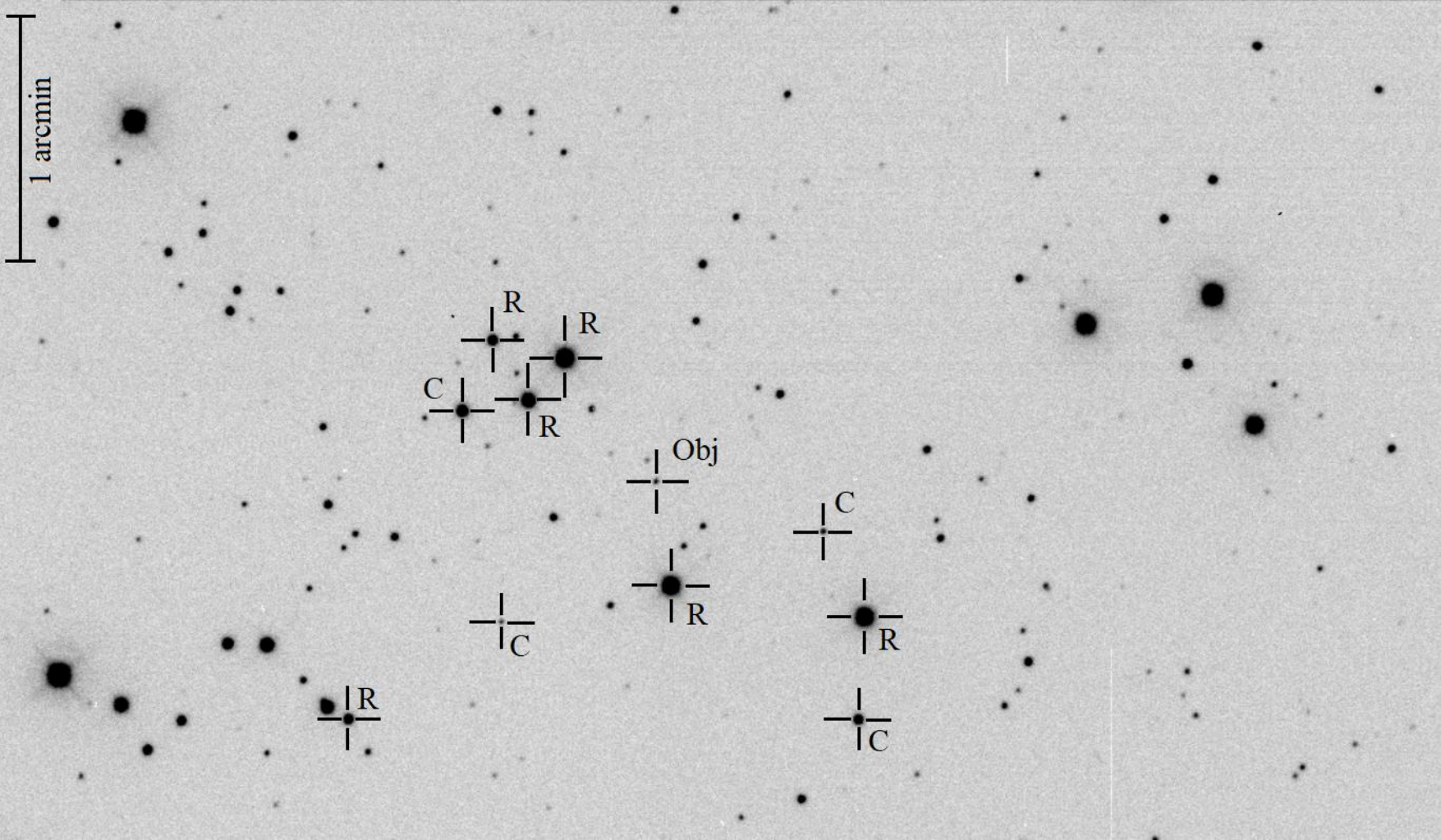}
    \caption{The \ZTF field. Obj - the object, R - standard stars, C - comparison stars. North is up, east to the left.}
    \label{Fig1}
\end{figure}

The brightness measurements were carried out using the method of aperture photometry in the $AstroImageJ$ software \citep{2017AJ....153...77C}. The $14^m-19^m$ in the $R_c$ filter were selected as comparison stars (see Fig. \ref{Fig1}).
\begin{table}
\begin{center}
\caption{Journal of observations. 2.5-m telescope of CMO SAI MSU, filter $R_c$}
\vspace{2mm}
\label{tab:log}
\begin{adjustbox}{width=0.5\textwidth}
\begin{tabular}{llc}
\hline\noalign{\smallskip}
JD interval & Date & Number of\\
& &CCD frames\\
\hline\noalign{\smallskip}

2460416.1972 –- .2767 & 15/04/2024 & 392 \\
2460417.2089 –- .2696 & 16/04/2024 & 319 \\
2460584.4216~--~.5475 & 30/09-1/10/2024 & 916 \\
2460624.3809 –- .5053 & 9-10/11/2024 & 906 \\
2460670.1992 –- .3412 & 25/12/2024 & 901 \\
2460704.2232 –- .3571 & 28/01/2025 & 963 \\
2460941.4437 –- .5551 & 22-23/09/2025 & 809 \\
2460952.5255 –- .5914 & 4/10/2025 & 480 \\
\noalign{\smallskip}\hline
\end{tabular}
\end{adjustbox}
\end{center}
\end{table}
\begin{table}
\caption{
Parameters of 4k$\times$4k CCD-camera on 2.5-m CMO SAI MSU telescope
}
\label{tab:ccd_char}
\centering
\vspace{2mm}
\begin{adjustbox}{width=0.5\textwidth}
\begin{tabular}{l|c}
\hline
Parameter &   \\
\hline
Type        &   CCD \\
Formate     & $4096\times4096$ \\
Pixel sixe, mcm, & 15$\times$15        \\
Angular scale, $^{\prime\prime}$/pxl& 0.154        \\
ACD bit rate, bit    &   18  \\
Transformation coefficient \\
(1 channel, bin 1), $e^{-}$/ADU&  0.251  \\
Transformation coefficient \\
(1 channel, bin 2), $e^{-}$/ADU&  0.243  \\
Transformation coefficient \\
(4 channels, bin 2), $e^{-}$/ADU&  0.244, 0.247, 0.247, 0.246  \\
reading noise rms\\ (1 channel, bin 1), $e^{-}$&   4.57  \\
reading noise rms \\(1 channel, bin 2), $e^{-}$&   4.68  \\
reading noise rms\\(4 channels, bin 2), $e^{-}$&   4.64, 5.52, 5.94, 5.09  \\
Dark current at $-85^{\circ
}$C \\(1 channel, bin 1), $e^{-}$/pxl/s &   0.0022  \\
Dark current at $-85^{\circ
}$C  \\(1 channel, bin 2), $e^{-}$/pxl/s &   0.022  \\
Dark current at $-85^{\circ
}$C \\(4 channels, bin 2), $e^{-}$/pxl/s &   0.025  \\
Reading time into 1 channel,\\
bin 1, s&  18.7  \\
Reading time into 1 channel, \\bin 2, s&  5.6  \\
Reading time into 4 channels, \\bin 2, s&  1.4  \\

\hline
\end{tabular}
\end{adjustbox}
\end{table}



\section*{Characteristics of the CCD camera of the CMO SAI 2.5-m telescope}

In this work, we used a new photometric $4096\times4096$ CCD camera with a pixel size of $15\times15$~ microns, with back illumination with deep depletion technology and suppression of parasitic interference. A multilayer anti-reflection coating is applied to the sensitive surface of the CCD detector, providing a high quantum efficiency of more than 85\% in the range from 4000 to 8000\AA. The characteristics of the detector are listed in Table.~\ref{tab:ccd_char}. The detector is placed in a high-vacuum chamber, the entrance window is made of fused quartz without an anti-reflective coating to preserve a wide transmission range. External shutter, curtain, opening/closing time is $\approx.120$~ms. The camera is installed in the Cassegrain focus of the CMO SAI 2.5-m telescope. The camera's field of view is $10.5^{\prime}\times10.5^{\prime}$. The angular scale is $0.154^{\prime\prime}$/pcx. The conversion factor is 0.25~$e^{-}$/ADU. The reading speed is 1 MHz. The device has the ability to read into 1, 2 or 4 channels simultaneously. When read into 4 channels, the frame is divided into four quadrants read in parallel. We used the following combinations of reading modes:
\begin{itemize}
\item into 1 channel without binning,
\item into 1 channel with binning 2,
\item into 4 channels with 2 binning.
\end{itemize}


The CCD detector is cooled with a cascade of Peltier elements, heat is removed by a liquid circuit; the temperature of the coolant is maintained at 12${}^\circ\mathrm{C}$. The temperature of the CCD detector is set depending on the ambient conditions and the capabilities of the heat sink in the range from $-80$ to $-90{}^\circ\mathrm{C}$.


The CCD camera control software consists of a main control program written in C++ and a graphical interface implemented in PyQt5. These programs communicate with each other via a TCP/IP socket using a specially designed and implemented protocol. The main control program interacts with the detector, sets the necessary shooting parameters, receives and records frames in FITS files, and communicates with the telescope's control program and the filter wheel control program. Using a graphic interface, the observer has the opportunity to focus, upload or create observation scenarios for objects (i.e., the necessary combinations of exposition parameters - filter, number of reading channels, binning, number of frames), as well as view the resulting images.

\begin{figure}
    \centering
    \includegraphics[width=0.8\linewidth]{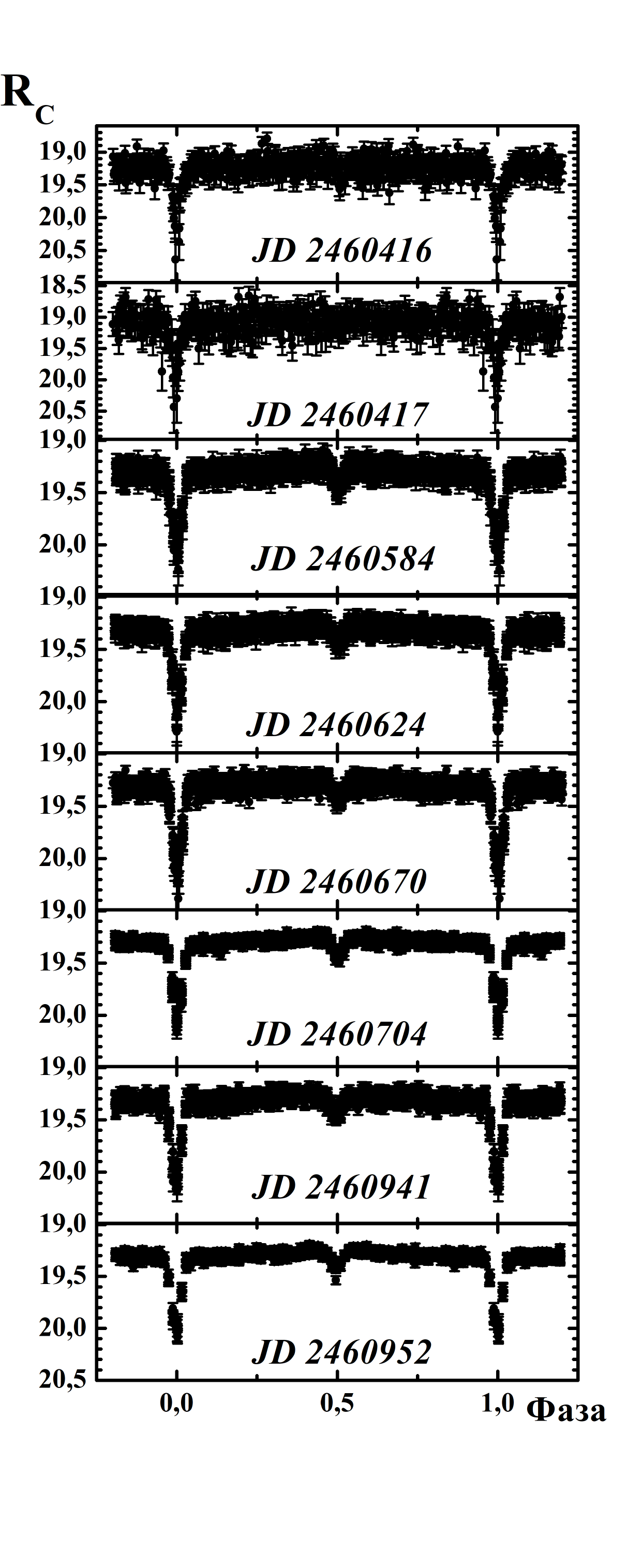}
    \caption{Phase light curves if \ZTF in individual nights calculated using ephemerid (1).}
    \label{Fig2}
\end{figure}

\begin{figure}
    \centering
    \includegraphics[width=0.9\linewidth]{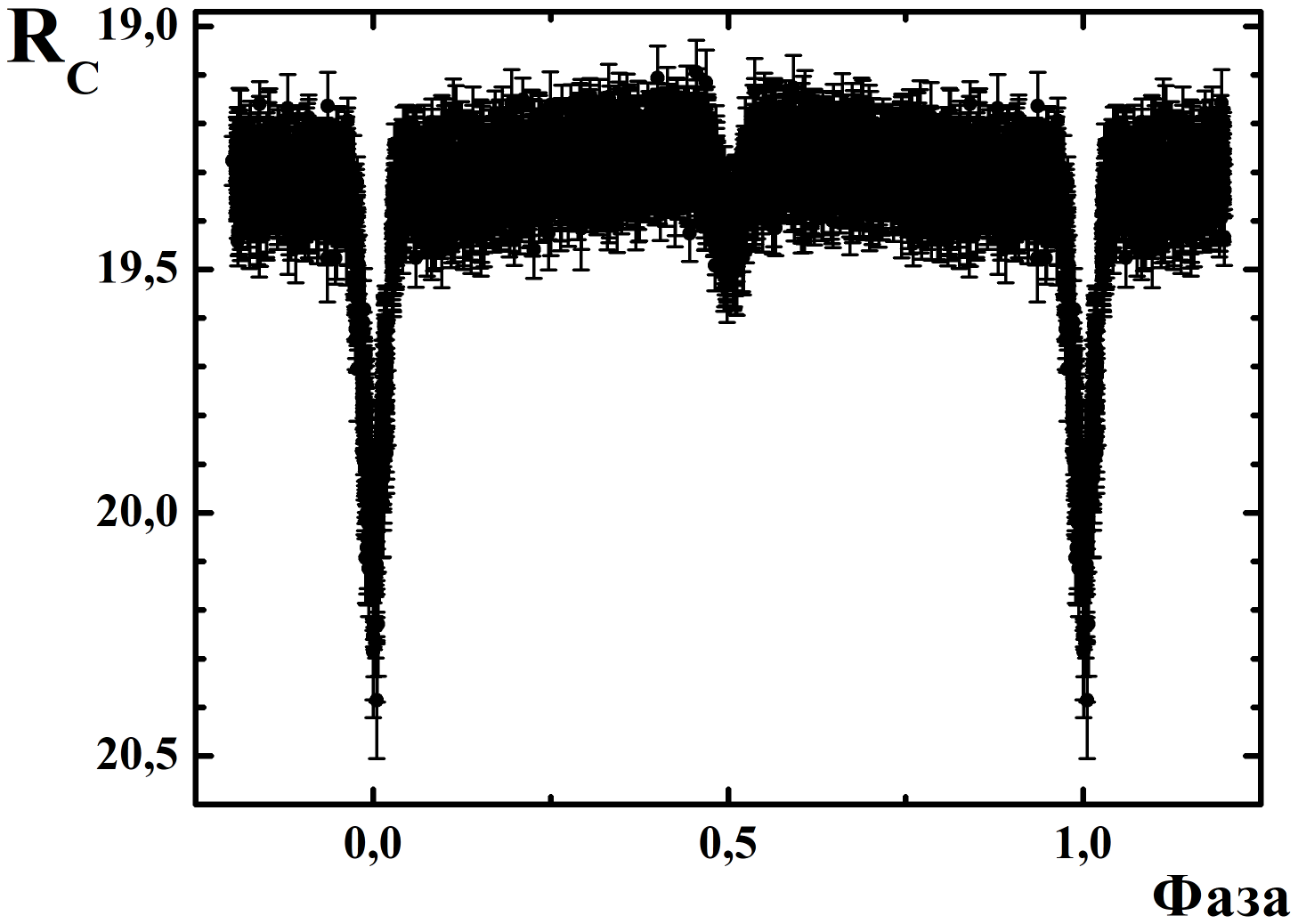}
    \caption{Phase light curve of \ZTF calculated from six observational nights (JD 2460584-60952).}
    \label{Fig3}
\end{figure}

\section{Analysis of the \textit{O-C} diagram}


For the most accurate determination of the moments of the main eclipse minima \ZTF we applied the Hertzsprung method \citeyearpar{1919AN....210...17H} algorithmized by Berdnikov \citeyearpar{1992SvAL...18..207B}. The moments of the main minima, together with the corresponding values of the residual deviations O–C, are shown in Table ~\ref{tab:oc}. The \textit{O-C} residuals are calculated using linear elements of the brightness  variation:
\begin{equation}
\mathrm{BJD\,Min} = 2 460 670.268615 + 0.0100301160 \times E.
\end{equation}

\begin{table}
\begin{center}
\caption{Moments of main minima of \ZTF}
\vspace{2mm}
\label{tab:oc}
\begin{adjustbox}{width=0.5\textwidth}
\begin{tabular}{llcl}
\hline\noalign{\smallskip}
Barycentric JD	& Error, day & \it{O-C}, day & Epoch\\
\hline\noalign{\smallskip}

2460416.2358360	& 0.0000301	& -0.00003107 & -25327\\
2460417.2388316	& 0.0000429	& -0.00004706 & -25227\\
2460584.4810329 & 0.0000075	& -0.00001094 & -8553 \\
2460624.4410153	& 0.0000069	& 0.00000029 & -4569\\
2460670.2686151	& 0.0000074	& 0.00000015 & 0\\
2460704.2907625	& 0.0000046	& -0.00000594 & 3392\\
2460941.5029679	& 0.0000076	& -0.00004393 & 27042\\
2460952.5561528	& 0.0000069	& -0.00004686 & 28144\\
\noalign{\smallskip}\hline
\end{tabular}
\end{adjustbox}
\end{center}
\end{table}


The phase light curves for individual observational nights convolved with elements (1) are shown in Fig.~\ref{Fig2}. The weather conditions on the first and second nights of observations (April 15 and 16, 2024) were not ideal, which affected the accuracy of photometry. The phase light curve, constructed separately for six nights with photometric weather conditions (November 9/10, September 30/October 1 and December 25, 2024, January 28, September 22/23 and October 4, 2025), is shown in Fig.~\ref{Fig3}.

\begin{figure}
    \centering
\includegraphics[width=0.9\linewidth]{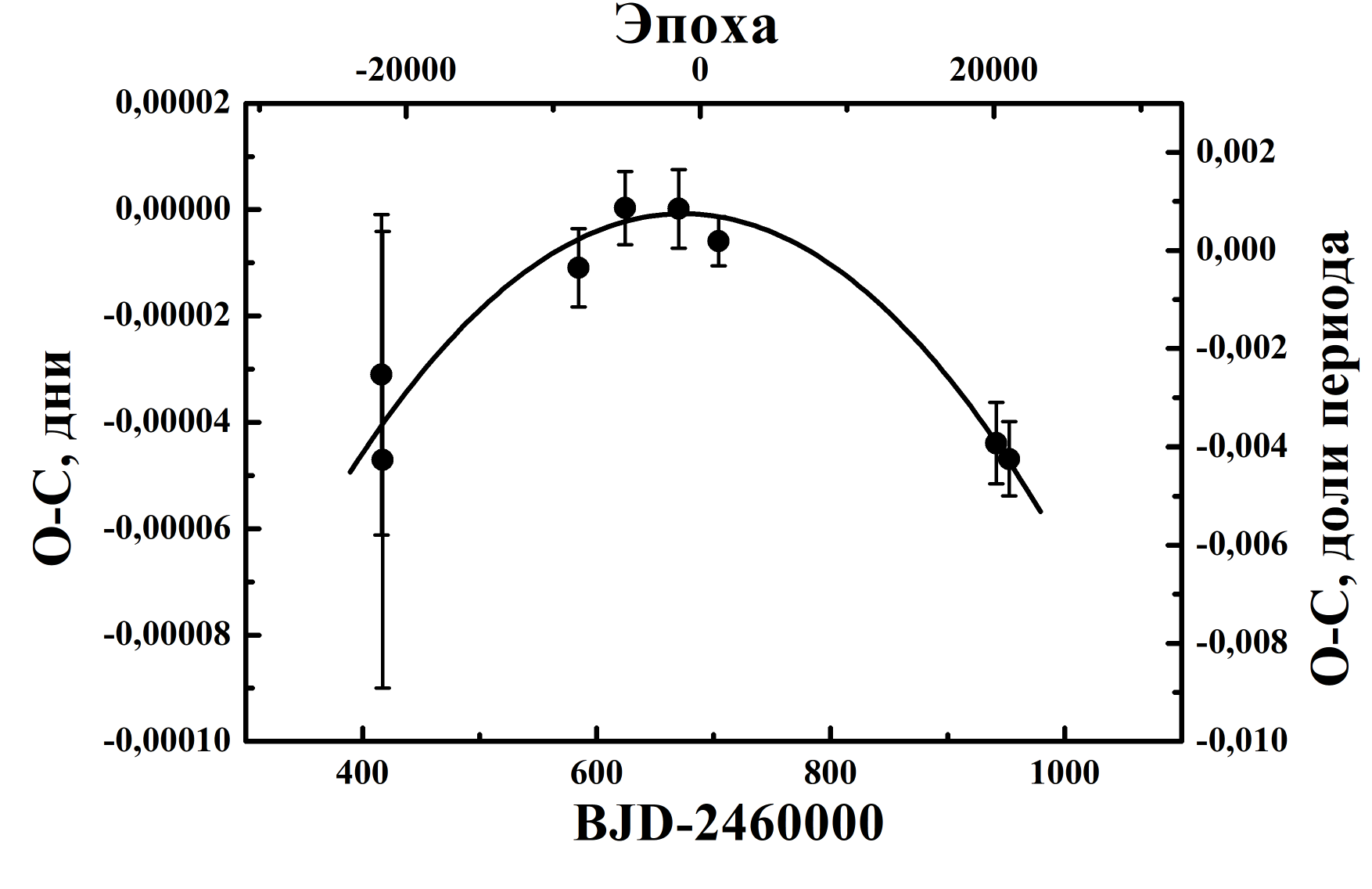}
    \caption{The O–C diagram of \ZTF relative to the linear ephemerid (1). The solid curve corresponds to quadratic elements (2).}
    \label{Fig4}
\end{figure}

Fig.~\ref{Fig4} shows the residual deviations of \textit{O-C} relative to the linear elements of the  brightness change (1).  Fig.~\ref{Fig4} suggests that it is impossible to describe observations using linear elements of brightness variation. The plot of the O-C residuals clearly demonstrates a quadratic term corresponding to a linear decrease in the period with time.:
\begin{equation}
 \begin{split}
\mathrm{BJD\,min_I} = 2 460 670.268615 (\pm{0.000008})\\
\\+ 0.0100301160 (\pm{0.0000000002}) \times E\\
\\-5.821 (\pm{1.119}) × 10^{-14} \times {E^2}.
 \end{split}
\end{equation}


The quadratic elements of the brightness change (2) imply a linear decrease in the orbital period of the binary system at a rate of $dP/dt=-(1.16\pm 0.22)\times 10^{-11}$ s/s. The significance of the quadratic term in the elements (2) obviously exceeds 3$\sigma$. An important advantage of our analysis is the uniformity of the data. All observations were performed on the same telescope with the same CCD matrix, processed using a single technique, and uniformly reduced to the time moments corresponding to the Solar system barycenter. This ensures high reliability of the result obtained.

\section{Orbital period decay of \ZTF}

The measured value of the decrease rate of the orbital period of the ultrashort-period binary system of two white dwarfs \ZTF $dP/dt=-(1.16\pm 0.22)\times 10^{-11}$ s/s allows us to determine the value of its chirp mass $\mathcal{M}$ from the quadrupole formula for GW radiation:
\begin{equation}
    \frac{dP}{dt}=-\frac{96}{5}(2\pi)^{8/3}\left(\frac{G\mathcal{M}}{c^3}\right)^{5/3}P^{-5/3}\,.
\end{equation}
The resulting value of $\mathcal{M}=0.434\pm 0.05 M_\odot$ turns out to be higher than the chirp mass calculated from the component masses obtained from spectroscopic observations of $M_1=0.45\pm 0.05 M_\odot,\, M_2=0.32\pm0.03 M_\odot$ \citep{2020ApJ...905...32B}: $ \mathcal{M}_{sp}=0.33 \pm 0.03 M_\odot$.

For the eclipsing binary white dwarf \ZTF, the orbital inclination is $i\simeq 85^\circ$\citep{2020ApJ...905...32B}, therefore  in the quadrupole approximation the GW polarization  $h_+\sim\mathcal{A}(1+\cos^2i)$ 
mainly contributes to the GW signal amplitude
(the polarization $h_\times\sim 2\mathcal{A}\cos i$ and is negligible):  $h_+\approx\mathcal{A}= 2(G\mathcal{M})^{5/3}(\pi f_{GW})^{2/3}/c^4d$ ($d$ is the distance to the source, $f_{GW}=2/P$ is the fundamental harmonic of the GW frequency for a circular orbit).  Since $\mathcal{M}^{5/3}\sim -\dot P P^{5/3}$ (see (3)), then the estimate of the GW signal amplitude from \ZTF is read off directly from the measurement of $\dot P$:
\begin{equation}
    \mathcal{A}\approx\frac{10}{96}\frac{(-P\dot P)}{(2\pi)^2}\frac{c}{d}\,.
\end{equation}
With a distance to \ZTF of $d\approx 1$ kpc we find $\mathcal{A}\approx 2.6\times 10^{-22}$. The signal-to-noise ratio during observational time $T$ for a source with a constant (or slightly variable during observations) frequency $f_0$ by a LISA-type space laser interferometer \citep{2019CQGra..36j5011R,2020PhRvD.102f3021H}
\begin{equation}
    \rho\approx \frac{\sqrt{2T}\mathcal{A}}{\sqrt{S_n(f_0)}}
\end{equation}
($\sqrt{S_n(f_0)}$ [Hz$^{-1/2}$] is the detector's noise sensitivity curve ). Because of formula (4) the signal-to-noise ratio depends directly on $\dot P$:
\begin{equation}
    \rho\sim \frac{\sqrt{T}}{d \sqrt{S_n(2/P)}}\left(-\dot PP\right)\,.
\end{equation}
As the obtained GW amplitude derived from the measurement of $\dot P$ in \ZTF is by almost 58\% higher than the estimate from the spectroscopic determination of the component masses, the expected signal-to-noise ratio when measuring the GW signal from this verification binary system by both the TianQin detector over 2.5 years of observations and the LISA detector over 5 years of observations (see Table 3 in \cite{2023ApJS..264...39R}) also increases by 58\%. This makes \ZTF the brightest verification binary source for LISA ($\rho\approx 119$) and TianQin ($\rho\approx 30$) laser interferometers (see Table \ref{tab:rho}).
\begin{table}
\begin{center}
\caption{Parameters of GW signal from \ZTF.}
\vspace{2mm}
\label{tab:rho}
\begin{adjustbox}{width=0.5\textwidth}
\begin{tabular}{rcc}
\hline\noalign{\smallskip}
Parameter & From spectroscopic data&
From $\dot P$ measurements\\ &\citep{2023ApJS..264...39R} & (the present paper)\\
\hline\noalign{\smallskip}
Chirp mass $\mathcal{M}$, $M_\odot$ & 0.33 & $0.434$\\
Amplitude $\mathcal{A}$, $\times 10^{-23}$ & 16.38 & 25.86\\ 
Signal-to-noise ratio $\rho$ \\
LISA (5 years) & 75.532 & 119\\
TianQin (2.5 years) & 18.915 & 30\\
\noalign{\smallskip}\hline
\end{tabular}
\end{adjustbox}
\end{center}
\end{table}

We also note that the measurement of $\dot P$ from a short-period verification binary, which leads to an independent estimate of the chirp mass, together with the future determination of the amplitude of the GW signal $\mathcal{A}$, will enable independent estimate of the distance to the source $d$.
The results obtained emphasize the paramount importance of determining the rate the orbital period decay of ultraclose binary systems -- verification binaries for space laser interferometers, which can be measured as a result of long-term photometric monitoring.

With the measured rate of decrease in the orbital period, the time to system coalescence of \ZTF according to the quadrupole formula is $t_0=\frac{96}{256}\left(\frac{P}{\dot P}\right)\approx 2.7\times 10^6$ years. Of course, a little earlier, a less massive component with a larger radius will fill its Roche lobe and the components may merge to form a rapidly rotating white dwarf with a mass of about $0.7 M_\odot$.

\section{Conclusion}

This paper presents the results of an analysis of photometric observations of the eclipsing binary white dwarf \ZTF with an ultrashort orbital period of $14.44$ min, conducted on the 2.5-m telescope of the CMO SAI MSU. The method of reduction of photometric measurements on the modified equipment of the CMO SAI MSU 2.5-m telescope is described in detail. Based on the results of photometric observations, we applied  the Hertzsprung method to determine the moments of the main minima of the eclipsing system and to construct the \textit{O-C} diagram. From the analysis of the \textit{O-C} diagram we calculated the quadratic elements of the brightness variations, indicating  decrease in the orbital period of the binary system at a rate of $dP/dt=-(1.16\pm 0.22)\times 10^{-11}$ s/s. Assuming the orbital period decay due to GW radiation, the chirp mass of the binary system in the quadrupole approximation is determined  $\mathcal{M}=0.434\pm 0.05 M_\odot$, which turns out to be higher by $\sim 30\%$ than the value obtained from the spectroscopic determination of the component masses in paper \citeyearpar{2020ApJ...905...32B}. 
The obtained value of the chirp mass of \ZTF from measurements of the rate its orbital period decay makes this system the brightest Galactic verification binary for LISA and TianQin space interferometers with a signal-to-noise ratio of $\approx 119$ and $\approx 30$ over 5 years and 2.5 years of observations, respectively.
The importance of direct determination of the the orbital period decrease rate of ultraclose binary systems -- potential Galactic verification sources for space laser interferometers -- is emphasized, since the expected signal-to-noise ratio when registering (almost) monochromatic GWs for such sources is directly proportional to $-\dot P P$.



Additional precision photometric observations of \ZTF will reduce the measurement errors of $\dot P$ and the output parameters of the GW signal from this binary white dwarf. 

\section*{Acknowledgements}

The work of S.V. Antipin, N.A. Maslennikova and K.A. Postnov was supported by the Russian Science Foundation grant 23-42-00055.
The work was performed using equipment purchased in the frame of the Lomonosov Moscow State University Development Program.

\section*{Data availability}
Data are available from the authors by request.





\bsp	
\label{lastpage}
\end{document}